\documentstyle[prd,aps,floats,twocolumn]{revtex}

\preprint{DAMTP-1999-31}
\date{Phys. Lett. B: Submitted 1/3/99, revised 1/6/99, accepted 6/6/99}
\tighten
\begin{document}
\draft
\def\sqr#1#2{{\vcenter{\hrule height.3pt
      \hbox{\vrule width.3pt height#2pt  \kern#1pt
         \vrule width.3pt}  \hrule height.3pt}}}
\def\square{\mathchoice{\sqr67\,}{\sqr67\,}\sqr{3}{3.5}\sqr{3}{3.5}}
\def\today{\ifcase\month\or
  January\or February\or March\or April\or May\or June\or
  July\or August\or September\or October\or November\or December\fi
  \space\number\day, \number\year}


\title{Does a varying speed of light solve the cosmological problems?}

\author{P. P. Avelino${}^{1}$\thanks{
Electronic address: pedro\,@\,astro.up.pt} 
and C. J. A. P. Martins${}^{2}$\thanks{Also at C.A.U.P.,
Rua das Estrelas s/n, 4150 Porto, Portugal.
Electronic address: C.J.A.P.Martins\,@\,damtp.cam.ac.uk}}

\address{${}^1$ Centro de Astrof\'{\i}sica, Universidade do Porto\\
Rua das Estrelas s/n, 4150 Porto, Portugal}

\address{${}^2$ Department of Applied Mathematics and Theoretical Physics\\
University of Cambridge, Silver Street, Cambridge CB3 9EW, U.K.}

\maketitle
\begin{abstract}

{We propose a new generalisation of general relativity which incorporates 
a variation in both the speed of light in vacuum (c) and the gravitational
constant (G) and which is both covariant and Lorentz invariant. We solve 
the generalised Einstein equations for Friedmann universes and show that 
arbitrary time-variations of c and G never lead to a solution
to the flatness, horizon or $\Lambda$ problems for a theory 
satisfying the strong energy condition. In order to do so, one
needs to construct a theory which does not reduce to the standard one 
for any choice of time, length and energy units. This can be achieved 
by breaking a number of invariance principles such as covariance and 
Lorentz invariance.}

\end{abstract} 
\pacs{PACS number(s): 98.80.Cq, 95.30.St}
\newpage

\section{Introduction}
\label{secintro} 

Inflationary models were originally proposed as a solution to some of the most
fundamental problems of the standard cosmological model namely the
horizon, flatness and monopole problems \cite{G,L1,AS,L2}. In the context 
of inflation the solution to these problems is achieved through a period of 
very rapid expansion induced by a huge vacuum energy.

Despite the lack of a single, well motivated particle physics model for
inflation, these models are highly successful in providing
solutions to such cosmological puzzles. Still, it is crucial to investigate if other scenarios could also solve some of these cosmological problems, or 
even others which inflationary models do not address (such as 
the cosmological constant problem). In particular, it is important to establish
what general conditions are required of a theory which is capable of providing
solutions to such problems.

There have been recent claims for a time-varying fine
structure constant \cite{WFCDB} detected by comparing quasar spectral lines in different multiplets. These possible 
variations in the dimensionless parameter $\alpha$ can be interpreted as 
variations in dimensional constants such as the electric charge, the Planck 
constant or the  speed of light in vacuum \cite{BE,B,BM}. 
Albrecht and Magueijo \cite{AM} have recently proposed a generalisation of
General  Relativity incorporating a possible change in the speed of light in
vacuum (c) and the gravitational constant (G)---see also \cite{B,M1,M2}. They have shown that 
their theory can solve many of the problems of the standard cosmological model including the horizon, flatness and cosmological constant problems, at the price
of breaking covariance and
Lorentz invariance. The also make the additional
{\it ex nihilio} assumption of minimal coupling at the level of
Einstein's equations.

Here we take a pedagogical look at this problem by asking if one can restore
some of the above principles and still obtain a theory which, {\it
prima facia}, provides credible solutions to the standard cosmological enigmas.
With this aim, we propose a new generalisation of General Relativity which
also allows for arbitrary changes in the speed of light, $c$, 
and the gravitational constant, $G$. Our theory is both covariant and Lorentz 
invariant and for $G/c^2 = {\rm constant}$ both mass and particle number are 
conserved. We solve the Einstein equations for
Friedmann universes and show that the solution to the flatness,
horizon or $\Lambda$ problems always requires similar conditions to the ones found in the context of the standard cosmological model. 

We therefore argue that a theory that reduces to General Relativity in
the appropriate limit and solves the horizon and flatness
problems of the standard cosmological
problems must either violate the strong energy condition (which is what
inflation does), Lorentz invariance or covariance. Stronger requirements are
needed in order to solve also the cosmological constant problem.
In a subsequent publication \cite{AM2} we shall show that our approach and that
of Albrecht and Magueijo \cite{AM} can be further distinguished by an
analysis of their corresponding structure formation scenarios.

\section{A variable speed of light theory}
\label{secopen}

Experiments can only measure 
dimensionless combinations of the fundamental parameters. This means that 
any evidence for variation in a dimensional parameter is dependent on
the choice of units in which it is measured. Hence, before
investigating the cosmological consequences of a variable speed of
light theory we must specify our choice of units. 

Here, we choose our unit of energy to be 
the Rydberg energy $E_R = m_e e^4/ 2(4\pi \epsilon_0
)^2\hbar^2$, our unit of length to be the Bohr radius ($a_0 =
4\pi\epsilon_0\hbar^2 / m_e e^2$) and our time unit to be $\Delta t = \hbar /
E_R$. Using these units a measure of the velocity of light will be 
a measure of the dimensionless quantity 
\begin{equation}\label{calpha}
{{c \hbar} \over {a_0  E_R}} = {{8 \pi \epsilon_0} \over \alpha}
\end{equation}
where $\alpha = e^2/ (\hbar c)$ is the fine structure constant. Hence, 
by choosing appropriate units we are able to interpret a variation in the 
fine structure constant $\alpha$ as being due to a change in the speed of 
light, c. It is possible to redefine our unit of time in such a way that 
$c$ and $e$ remain fixed while $\hbar$ varies proportionally to 
$\alpha^{-1}$ by making
\begin{equation}\label{newtime}
\Delta t' = \alpha \Delta t = {{\alpha \hbar} \over  E_R}.
\end{equation}
In this article we shall not address the problem 
of which mechanism could induce a change in $\alpha$ but we concentrate 
on the cosmological implications of such a variation. 

Given that c is a constant in these units,
we may specify our theory of gravity to be Einstein's General Relativity 
with a variable gravitational constant $G$. The Planck constant ($\hbar$) and 
the gravitational constant ($G$) will be a function of the space-time 
position. However, we will 
assume, for the sake of simplicity, that to zeroth order $\hbar$ and $G$ are 
functions 
of the cosmological time only. The theory specified in this 
way will clearly be Lorentz invariant. Our theory implies a modification of quantum mechanics in order to 
incorporate a variable $\hbar$ (as indeed do the theories discussed
in \cite{B,BM,AM}). However, for $E_R/ {\dot \hbar} << 1$ 
the variation of the Planck constant is very small on an atomic 
timescale. This means that quantum mechanical results for atomic behaviour 
will hold to a very good approximation with only a simple 
modification $\hbar \to \hbar (t)$. Nevertheless, we do expect that
such changes will have observational consequences (eg. for black body curves).
We will discuss this in more detail elsewhere, but here we simply point out
that this (and many other constraints) will force any significant changes in
the fundamental constants to happen very early in the history of the universe.

If we now switch to our original 
(and more natural) choice of units we will be left with a theory which 
has a variable $c$ but which is just analogous to General Relativity.
The mass of an electron is a constant in these units, and it 
will be implicitly assumed that in the absence of any interactions 
(including gravity) the average distance between particles will 
remain a constant. From now on we will stick to our original 
choice of time unit $\Delta t = \hbar / E_R$. In passing, we note that it
would also be possible,
by making appropriate changes of units, to interpret the variation of the
fine structure constant as a variation in the electric charge $e$ \cite{BE}.
This different choice of units would lead to a different interpretation of
the theory, but at the end of the day the physical consequences of the model would be the same.

In our model the Einstein equations take the usual form
\begin{equation}\label{einstein}
G_{\mu \nu} - g_{\mu \nu}\Lambda = {{8 \pi G} \over c^4} T_{\mu \nu},
\end{equation}
but now arbitrary variations in $c$ and
$G$ will be allowed. Contrary to Albrecht and Magueijo we do not assume
{\it ab initio} that variations in the 
speed of light do not introduce corrections to the curvature terms in the 
Einstein equations in the cosmological frame. In our model variations in the velocity of light, are always allowed to contribute to the curvature terms. 
These contributions are computed from the metric tensor in the usual way.
Note that our only assumption is that both $\alpha$ and (consequently) $c$
are a function of cosmological time $t$ only, to zeroth order. 
One can see that this is essentially similar to the much more familiar
assumption that both density and pressure are functions of the cosmological
time only, to zeroth order in the metric perturbations. Consequently, this
does not break the covariance of the theory.

With the line element
\begin{equation}\label{linel}
ds^2=a^2\left[c^2 dt^2-\frac{dr^2}{1-Kr^2}-r^2d\Omega_2^2\right] \, ,
\end{equation}
the Friedmann and Raychaudhuri equations in our theory are given by
\begin{eqnarray}
{\left({\dot a\over a}\right)}^2&=&{8\pi G\over 3}\rho -{Kc^2\over a^2}
\label{friedman1}\\
{\ddot a\over a}&=&-{4\pi G\over 3}{\left(\rho+3{p\over c^2}\right)}+
{\dot c\over c}{\dot a\over a}.
\label{friedman2}
\end{eqnarray}
Here, $\rho c^2$ and $p$ are the energy and
pressure densities,
$K=0,\pm 1$ and $G$ the curvature and the gravitational
constants, and the dot denotes a derivative with respect to proper time.
These can be combined into a conservation equation
\begin{equation}\label{friedman3}
d{{(G \rho a^{3 \gamma} c^{-2})}}/dt = 0,
\end{equation}
where $\gamma$ is defined by $p=(\gamma-1) \rho c^2$.
If the factor $G/c^2$ is a constant then the mass density 
($\rho$) is conserved. In general, however, the conserved quantity will not
be what one usually defines as `energy'. This is due to our particular
choices of `fundamental' units. Unlike in the theory proposed by Albrecht
and Magueijo the curvature of the universe does not explicitly appear
in the conservation equation. As we will show elsewhere \cite{AM2}, this
difference is crucial for the ensuing structure formation scenarios.

Note that it is easy to transform between our general coordinate
system as specified by the line element (\ref{linel}) and one in which
$c_0=1$ (we use the subscript zero to denote quantities measured in these
coordinates). The transformation rules are 

\begin{equation}\label{tr1}
dt_0=cdt \, ,
\end{equation}
\begin{equation}\label{tr2}
\rho_0=\rho c^2 \, ,
\end{equation}
\begin{equation}\label{tr3}
p_0=p \, ,
\end{equation}
\begin{equation}\label{tr4}
G_0=G/c^4 \, .
\end{equation}

It is then straightforward to check that, for example,
the Friedmann and Raychaudhuri equations (\ref{friedman1},\ref{friedman2}) transform in the correct way.

We note in passing that in our theory the Planck time given by 
$$
t_{\rm Pl}= {\sqrt {G \hbar \over c^5}}
$$
will be a variable for $G/c^5 \neq {\rm constant}$. This means that we 
may enter the quantum gravitational epoch sooner or later than in 
the standard cosmological scenario depending on the behaviour of both 
$c$ and $G$ at early times.

\section{The flatness, horizon and Lambda problems}
\label{flatness}

In order to solve the flatness problem the curvature term in equation 
(\ref{friedman1}) needs to be subdominant at late times. From the 
conservation equation we have that $G \rho \propto c^2 a^{-3 \gamma}$ and 
so equation (\ref{friedman1}) can be re-written as 
\begin{equation}\label{friedman4}
{\left({\dot a \over a}\right)}^2  =  C {c^2 \over a^{3 \gamma}} - 
K{c^2 \over a^2}.
\end{equation}
where $K$ and $C$ are constants.
We can see that the condition necessary for the the curvature term to 
be subdominant at large $a$ is 
\begin{equation}\label{conditionf}
\gamma \le 2/3.
\end{equation}
This is just the condition necessary to solve the flatness problem in the 
standard cosmological model. Consequently, no solution to the flatness 
problem arrises naturally in this model.

On the other hand, a solution to the horizon problem can be achieved by
having a period 
in the history of the universe in which the the scale factor can grow faster 
than the proper distance to the horizon,
\begin{equation}\label{horizons}
d_H = a(t)  \int_0^t {{c dt'} \over {a(t')}} .
\end{equation}
We can easily see that the scale factor can grow faster than $d_H$ only if
\begin{equation}\label{conditionh}
\gamma \le 2/3
\end{equation}
Hence, the condition necessary to solve the horizon problem is again
identical to the solution to the flatness problem and is no 
different from the one we obtain in the standard cosmological 
model: we must violate the strong energy condition.

Finally, a cosmological constant $\Lambda$ can be accounted for by including the cosmological constant 
mass density ($\rho_\Lambda \equiv \Lambda c^2/8\pi G$) in equations (\ref{friedman1}) and 
(\ref{friedman2}). In this case equation (\ref{friedman1}) becomes
\begin{equation}\label{friedman5}
{\left({\dot a\over a}\right)}^2 = C {c^2 \over a^{3 \gamma}} -{Kc^2\over a^2}
+{{\Lambda c^2} \over 3}.
\end{equation}
The condition necessary for the cosmological constant term in 
equation (\ref{friedman5}) to become negligible at late times is just  
\begin{equation}\label{conditionl}
\gamma \le 0.
\end{equation}
Again this is exactly the same condition we get in the standard cosmological 
model. We therefore conclude that our theory does not provide a solution to the standard cosmological problems. Note that our results do not depend on any
assumptions about the specific behaviour of $c$ and $G$. In particular, they
hold whether or not mass and particle number are conserved.

\section{Discussion and conclusions}
\label{conclusions}

In this article we have explicitly constructed a a generalisation
of General Relativity which is both
covariant and Lorentz invariant and obeys the strong energy condition, and
shown that in such a theory any
arbitrary time-like variations in $c$ and $G$ will not 
lead to a solution to some of the most important problems of the standard
cosmological model.
The main drawback of the theory we have constructed is that it is incomplete 
in the sense that a model for the dynamics of $c$ and $G$ is not presented.
Also, another outstanding issue which needs further discussion
is that of the possible observational
consequences of the required modifications to quantum mechanics.

From our above discussion, it is easy to see that the reason why
such a theory cannot solve the  standard cosmological enigmas is that one
can always find a choice of time unit in which this theory will be identical
(roughly speaking) to the standard cosmological model.
Furthermore, this should be true of {\em any} theory which is (a) covariant and
Lorentz invariant, (b) reduces to GR in the appropriate limit, and (c) obeys
the strong energy condition. The first two conditions ensure that there will
be a choice of units such that the theory in question will reduce to the
standard one, and then the condition required to solve the cosmological
problems should be that (c) is violated. Note that this argument is not strictly
a proof, since we have not provided a general way to find the required
choice of units. However, we believe that it is physically clear from the
discussion above (and in section \ref{secopen}) that such a choice of
units should exist.

On the other hand, the postulates of the theory of Albrecht and
Magueijo \cite{AM}, when translated in the above language, correspond to the
assumption that {\it one can not find any choice of time unit in which
the theory reduces to the standard one}. In their theory this is achieved by 
breaking covariance and Lorentz invariance.

Hence the above discussion leads us to conclude that to solve the standard
cosmological problems in a theory which reduces to General Relativity
(possibly after some appropriate changes in the `fundamental' units) one must
either violate the strong energy condition, Lorentz
invariance or covariance. Of course the above condition is necessary but not
sufficient. Inflation is an obvious example of a theory which violates the
first of the above principles.

A theory such as the one proposed by Albrecht and Magueijo \cite{AM}, on the
other hand, can work because it violates the latter two. In this context, the
fact that such a theory has a variable speed of light is, we think,  only a
minor `side effect' of their postulates, and other possibilities would do just
as well (eg, a varying electric charge \cite{BM}, etc). Furthermore, we anticipate that it should be possible to construct theories which break covariance and Lorentz
invariance, have `constant constants' and still can solve the horizon and
flatness problems, although one might also expect such theories to be even
more contrived than the ones discussed here.

In a subsequent paper \cite{AM2}, we will show that the approach of Albrecht
and Magueijo \cite{AM} has some other difficulties, in particular at the level
of structure formation scenarios. Some of these are due to the fact that there
are varying constants (and will therefore also appear in the theory we have
proposed in the present paper), but others are specifically due to the way
in which Lorentz invariance and covariance are broken in their approach.
This implies that it is not clear that this approach can be a viable
alternative to inflation. It is therefore interesting to ask if there
is any other paradigm, apart from these two, that can provide analogous
solutions to the cosmological enigmas.

\acknowledgements

We would like to thank Paulo Carvalho and Paulo Macedo for 
enlightening discussions.
P.P.A. is funded by JNICT (Portugal) under
`Programa PRAXIS XXI' (grant no. PRAXIS XXI/BPD/9901/96).
C.M. is funded by JNICT (Portugal) under
`Programa PRAXIS XXI' (grant no. PRAXIS XXI/BPD/11769/97).



\begin{references}
\bibitem{G}
A. Guth, {\it Phys. Rev. } {\bf D23}, 347 (1991).
\bibitem{L1}
A. Linde, {\it Phys. Lett. } {\bf B108}, 1220 (1982).
\bibitem{AS}
A. Albrecht \& P. Steinhardt, {\it Phys. Rev. Lett. } {\bf 48}, 1220 (1982).
\bibitem{L2}
A. Linde, {\it Phys. Lett. } {\bf B129}, 177 (1983).
\bibitem{WFCDB}
J. K. Webb, V. V. Flambaum, C. W. Churchill, M. J. Drinkwater \& J. D. Barrow,
{\it Phys. Rev. Lett. } {\bf 82}, 884 (1999)
\bibitem{BE}
J. D. Bekenstein, {\it Phys. Rev. } {\bf D25}, 1527 (1982).
\bibitem{B}
J. D. Barrow, {\it Phys. Rev. } {\bf D59}, 043515 (1999).
\bibitem{BM}
J. D. Barrow \& J. Magueijo, {\it Phys. Lett. } {\bf B443}, 104 (1998).
\bibitem{AM}
A. Albrecht \& J. Magueijo, {\it Phys. Rev. } {\bf D59}, 043516 (1999).
\bibitem{M1}
J. W. Moffat, {\it Int. J. Mod. Phys. } {\bf D2}, 351 (1992).
\bibitem{M2}
J. W. Moffat, {\it astro-ph/9811390} (1998).
\bibitem{AM2}
P. P. Avelino \& C. J. A. P. Martins, in preparation (1999).
\end{references}
\end{document}